\documentclass[12pt,preprint]{aastex}

\shorttitle{Magnetic Accretion Disk Spectra}
\shortauthors{Blaes et al.}

\begin{document}


\title{Magnetic Pressure Support and Accretion Disk Spectra}


\author{Omer M. Blaes and Shane W. Davis}
\affil{Department of Physics, University of California,
    Santa Barbara, CA 93106}
\email{blaes@physics.ucsb.edu, swd@physics.ucsb.edu}

\author{Shigenobu Hirose}
\affil{Earth Simulator Center, JAMSTEC, 3173-25 Showa-machi, Kanazawa-ku,
Yokohama, Kanagawa 236-0001, Japan}
\email{shirose@jamstec.go.jp}

\author{Julian H. Krolik}
\affil{Department of Physics and Astronomy, Johns Hopkins University,
Baltimore, MD 21218}
\email{jhk@pha.jhu.edu}

\and

\author{James M. Stone}
\affil{Department of Astrophysical Sciences, Princeton University,
Princeton, NJ 08544}
\email{jstone@astro.princeton.edu}


\begin{abstract}
Stellar atmosphere models of ionized accretion disks have generally neglected
the contribution of magnetic fields to the vertical hydrostatic support,
although magnetic fields are widely believed to play a critical role in
the transport of angular momentum.  Simulations of magnetorotational turbulence
in a vertically stratified shearing box geometry show that magnetic pressure
support can be dominant in the upper layers of the disk.  We present
calculations of accretion disk spectra that include this magnetic pressure
support, as well as a vertical dissipation profile based on simulation.
Magnetic pressure support generically produces a more vertically extended
disk atmosphere with a larger density scale height.  This acts to harden the
spectrum compared to models that neglect magnetic pressure support.  We
estimate the significance of this effect on disk-integrated spectra by
calculating an illustrative disk model for a stellar mass black hole, assuming
that similar magnetic pressure support exists at all radii.
\end{abstract}




\section{Introduction}

Almost without exception, models of geometrically thin accretion disks that
are used to predict radiation spectra in order to compare with observation
are based on the alpha prescription introduced by \citet{sha73}.  
This prescription enables one to compute the radial distribution of surface
density in the disk.  When this is combined with additional assumptions about
the vertical distribution of dissipation, complete models
of the vertical structure at each radius and the local radiation spectrum
can be computed.

In black hole accretion disks, for example, the most detailed of such models
so far are those of \citet{hub01} for supermassive black holes, \citet{dav05}
for stellar mass black holes, and \cite{hui05} for intermediate mass black
holes.  The models fully account for relativistic
effects, and include a detailed non-LTE treatment of level populations of
hydrogen, helium, and abundant metals.  Continuum opacities due to bound-free
and free-free transitions, as well as Comptonization, are included.  In spite
of this level of sophistication, the models are limited by other
assumptions:  the disk is assumed to be stationary, the alpha-prescription is
used to calulate the radial distribution of surface density, a no-torque inner
boundary condition is assumed, the vertical structure at each radius is
assumed to depend only on height and is symmetric about the midplane, the
vertical distribution of dissipation per unit mass is assumed to be
constant, and the disk is assumed to be supported vertically against the
tidal field of the black hole by just gas and radiation pressure.

It is now widely believed that the anomalous stress in accretion disks is
a result of sustained magnetohydrodynamical turbulence which is akin to
that seen in the nonlinear development of the magnetorotational instability,
or MRI \citep{bal98}.  In principle, shearing box simulations of this
turbulence which incorporate the vertical tidal field of the central mass
\citep{bra95, sto96, mil00} can be used to build models of the vertical
structure of accretion disks that are free of the ad hoc assumptions that
plague existing models.  The most recent of these simulations
\citep{tur04,hir05} are particularly useful for this purpose as they evolve
the equations of radiation magnetohydrodynamics and have self-consistent
thermodynamics.  While much of the dissipation of the turbulence in these
simulations is numerical and happens on the grid scale, one hopes that
this still mocks up the microscopic dissipation occuring at the smallest
scales of the turbulent cascade, and in any case provides a better handle
on the true spatial distribution of dissipation than the ad hoc assumptions
used in alpha-disk models.  Moreover, in radiation pressure dominated
simulations of MRI turbulence, dissipation of compressive motions by photon
diffusion is fully resolved \citep{tur03}.

\citet{dav05} used the time and horizontally-averaged vertical dissipation
profile from the simulation of \citet{tur04} to compute the vertical structure
and spectrum of an annulus, and compared it to a model based on the
usual ad hoc assumption of constant dissipation per unit mass.  While the
resulting profiles of density and temperature were very different, the
emergent spectrum was very similar.  In fact, \citet{dav05} found
that the locally emergent spectra of individual annuli were remarkably
robust to different vertical dissipation profiles and to different overall
stress prescriptions, provided most of the dissipation occurs deeper than
the effective photosphere.  Part of this robustness stems from the fact that
the radiative flux becomes constant above the dissipation region, and so
the radiative force per unit mass also becomes constant if the flux mean
opacity is dominated by electron scattering in the surface layers.  Because
the tidal gravity of the central mass increases with height, gas pressure
gradients must therefore compensate, and this results in very steep density
gradients near the photosphere.  The density and temperature at the effective
photosphere then become roughly independent of the details of the vertical
structure, resulting in a robust emergent spectrum.

In addition to a markedly different dissipation profile, vertically
stratified MRI simulations also show that the surface layers of a disk
will be {\it magnetically} supported.  This additional support removes
the need for steep density gradients, and in fact results in a much more
extended atmosphere with a larger density scale height.  As qualitatively
suggested by \citet{hir05}, magnetic pressure
support could therefore dramatically alter the emergent spectrum, much
more than alterations of the dissipation profile deeper than the effective
photosphere.  It is this new effect arising from MRI simulations that we
investigate in this paper.  

In section 2, we discuss the time and horizontally-averaged dissipation
and magnetic pressure profiles that we computed from the simulations.
We then used these profiles within the stellar atmosphere code TLUSTY
\citep{hub88,hub95} to compute the vertical structure and emergent spectrum of
individual annuli.  We present the results of these calculations in section 3,
including an illustrative whole disk spectrum for an accreting stellar mass
black hole.  We then summarize our conclusions in section 4.

\section{Vertical profiles of magnetic pressure and dissipation}

\citet{hir05} started with an initial condition corresponding to an annulus
at 300~$GM/c^2$ around a black hole of mass $M=6.62$~M$_\odot$.  The surface
density
of the annulus matched that of a Shakura-Sunyaev disk with $\alpha=0.02$ and
a luminosity equal to 0.066 times the Eddington rate, assuming 10 percent
radiative efficiency.  After including a weak magnetic field with no net
poloidal flux, the MRI grows and produces turbulence which saturates
by about ten orbits, and a rough balance between heating and radiative
cooling is thereafter established until a numerical problem at the boundaries
interferes at about 60 orbits.  At any specific time, thermal balance is not
exact, as there are significant fluctuations in both heating and radiative
cooling.  In addition, significant asymmetries above and below
the midplane exist, and the flux emerging from one side can exceed that
from the other by factors of 2-3 over time scales as long as 5-10 orbits.
Spatial variations in the horizontal direction come and go as well.

It would be interesting in future to compute the instantaneous emergent
spectrum from the annulus using three dimensional, time and frequency-dependent
radiative transfer, but our scope here is much less ambitious.  We simply
wish to examine the overall effects of the dissipation and magnetic
pressure profiles on the time-averaged spectrum.  We therefore horizontally
averaged all fluid quantities in the simulation at every time step, and then
computed a time average from 10 to 60 orbits of these variables at every
height.  [We tried two types of averages of quantities
related to energy and pressure (e.g flux, pressure gradients, dissipation
rate):  a straight time average and a time average weighted by the total
instantaneous magnetic plus thermal energy in the annulus.  Both approaches
gave very nearly identical results.] The time-averaged profiles still had
small asymmetries about the midplane (as much as 20 percent in the case of the
magnetic pressure gradients), so as a final step, we averaged all the vertical
profiles above and below the disk midplane.  The surface density and emergent
flux from the resulting averaged structure still matches that of a
Shakura-Sunyaev annulus at a radius of $300 GM/c^2$ around a 6.62~M$_\odot$
black hole, now accreting at $1.1\times10^{18}$~g/s (corresponding to a
luminosity of 0.095 in units of Eddington for 10 percent radiative efficiency)
and with $\alpha=0.016$.

\begin{figure}
\plotone{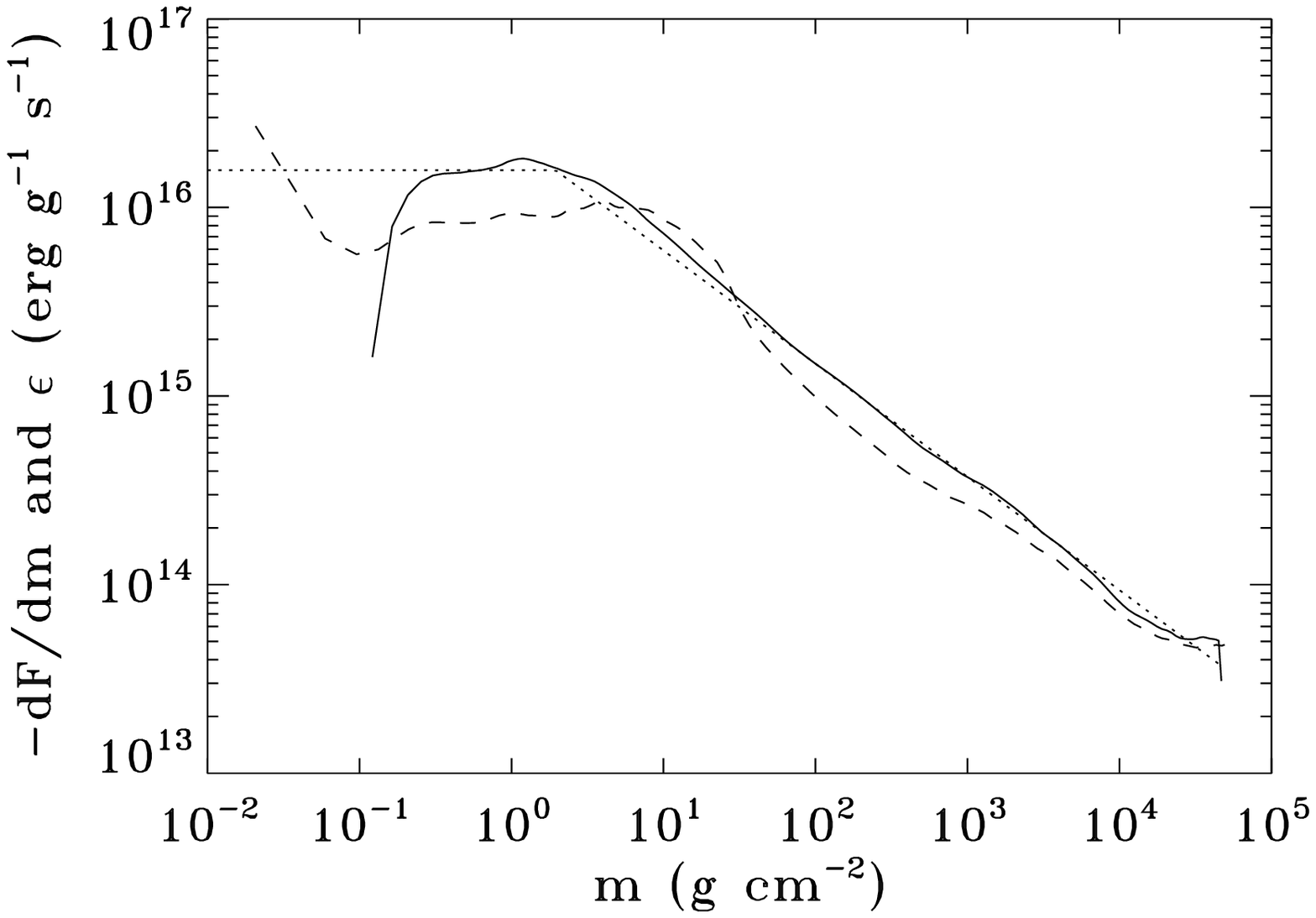}
\caption{Time and horizontally-averaged profiles of the vertical flux gradient
$\rho^{-1}dF/dz=-dF/dm$ (solid) and energy dissipation rate
per unit mass $\epsilon$ (dashed) as functions of column mass density $m$,
from the \citet{hir05} simulation.  The dotted curve is the broken power
law fit from equation (\ref{eqdfdmfit})  that we use in our spectral
computation.
\label{figdfdm}}
\end{figure}

Figure \ref{figdfdm} depicts the average energy dissipation per
unit mass $\epsilon$, together with the average vertical radiative
flux gradient divided by density: $\rho^{-1}dF/dz=-dF/dm$, as a function
of column mass density $m$ measured from the surface inward.  In a
stationary, one-dimensional structure,
radiative equilibrium would require these two profiles to be identical,
but the average profiles from the simulation are not quite in such
agreement.

Following \citet{hub98}, we fit the profile of $dF/dm$ with a broken power
law,
\begin{equation}
{dF\over dm}=-\left({F(0)\over m_0}\right){(\zeta_0+1)(\zeta_1+1)\over
(\zeta_0-\zeta_1)(m_d/m_0)^{\zeta_0+1}+\zeta_1+1}
\cases{(m/m_0)^{\zeta_1}(m_d/m_0)^{\zeta_0-\zeta_1},&if $m<m_d$;\cr
       (m/m_0)^{\zeta_0},&if $m>m_d$.\cr}
\label{eqdfdmfit}
\end{equation}
Here $F(0)=4.37\times10^{18}$~erg~cm$^{-2}$~s$^{-1}$ is the average flux
emerging from one face of the annulus
and $m_0=4.88\times10^4$~g~cm$^{-2}$ is the midplane column density of the
annulus (i.e. half the total surface mass density).
The dimensionless parameters for this fit were $\zeta_0=-0.6$, $\zeta_1=0$,
and $(m_d/m_0)=4\times10^{-5}$, and we compare the fit to the simulation
data in Figure \ref{figdfdm}.  Both the derivative $dF/dm$ from equation
(\ref{eqdfdmfit}) and its analytic integral $F(m)$ are used in constructing
the atmosphere model, because the radiative equilibrium equation in the
code TLUSTY is used in both integral and differential form \citep{hub95}.

\begin{figure}
\plotone{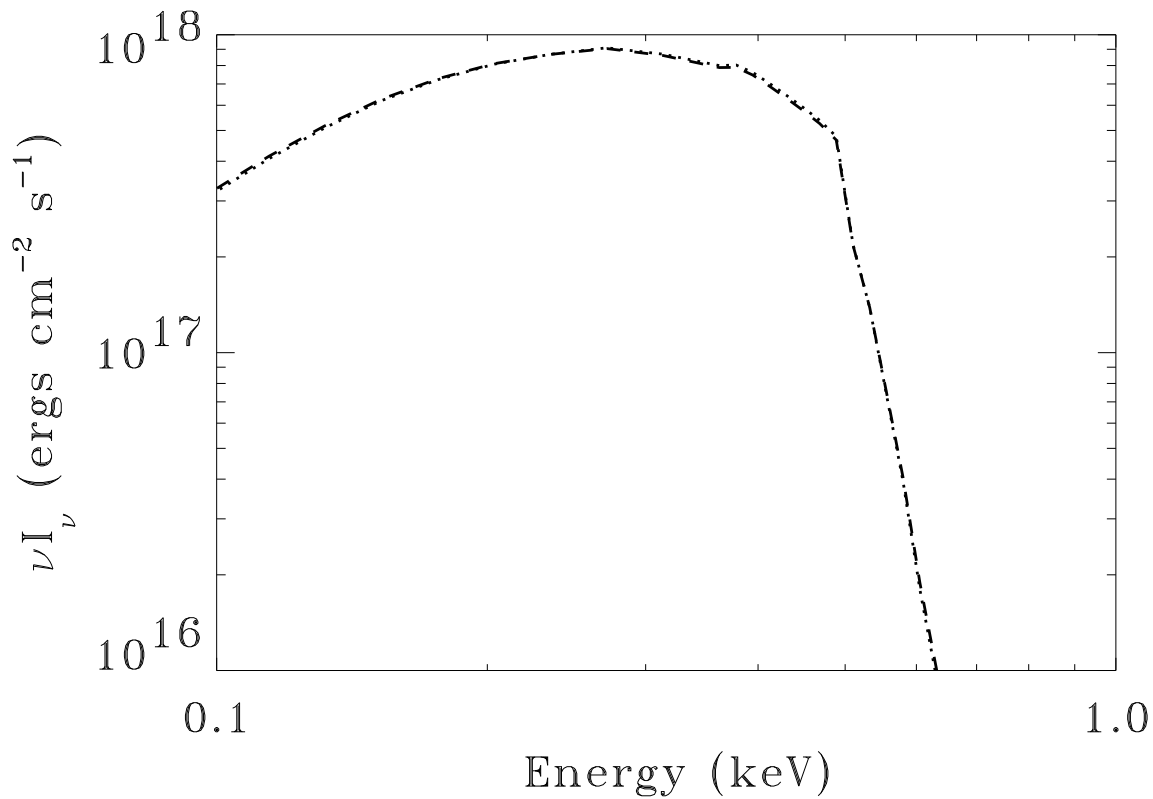}
\caption{Local emergent spectrum in the fluid rest frame from the disk annulus
viewed at an inclination angle of $55^\circ$ from the disk normal.  The
dotted curve is based on the standard assumption of constant dissipation per
unit mass, while the dashed curve uses the vertical dissipation profile of
equation (\ref{eqdfdmfit}) based on simulation.  Magnetic pressure support is
neglected in both cases shown here.  \label{figdissspec}}
\end{figure}

Note that $|m dF/dm|\propto m^{0.4}$ at high column densities, so that most
of the integrated dissipation in the annulus occurs at high column density
and low altitude, as \citet{hir05} point out.  While Figure \ref{figdfdm}
demonstrates that the highest values of dissipation per unit mass actually
occur at high altitudes and low column densities, this does not significantly
affect the emergent spectrum.  Figure \ref{figdissspec} compares the
emergent spectrum assuming a constant dissipation per unit mass with the
dissipation profile of equation (\ref{eqdfdmfit}).  The results are virtually
identical, although the spectrum based on equation (\ref{eqdfdmfit}) actually
differs in having a faint, hard tail at high energies.  This is not shown in
Figure \ref{figdissspec} as it starts too far into the tail of the spectrum.
The hard tail results from the fact that the new dissipation profile produces
a high temperature region at very low column densities, as shown in Figure
\ref{figpmagvertstruc} below.  In any case, provided most of the dissipation
occurs deeper than the effective photosphere, the resulting spectrum near the
spectral peak does not appear to depend sensitively on how the dissipation is
distributed.  This is perhaps not surprising, and \citet{dav05} found
very similar results when using the
dissipation profile computed by \citet{tur04}, although the emergent spectra
were not quite as close as here.

\begin{figure}
\plotone{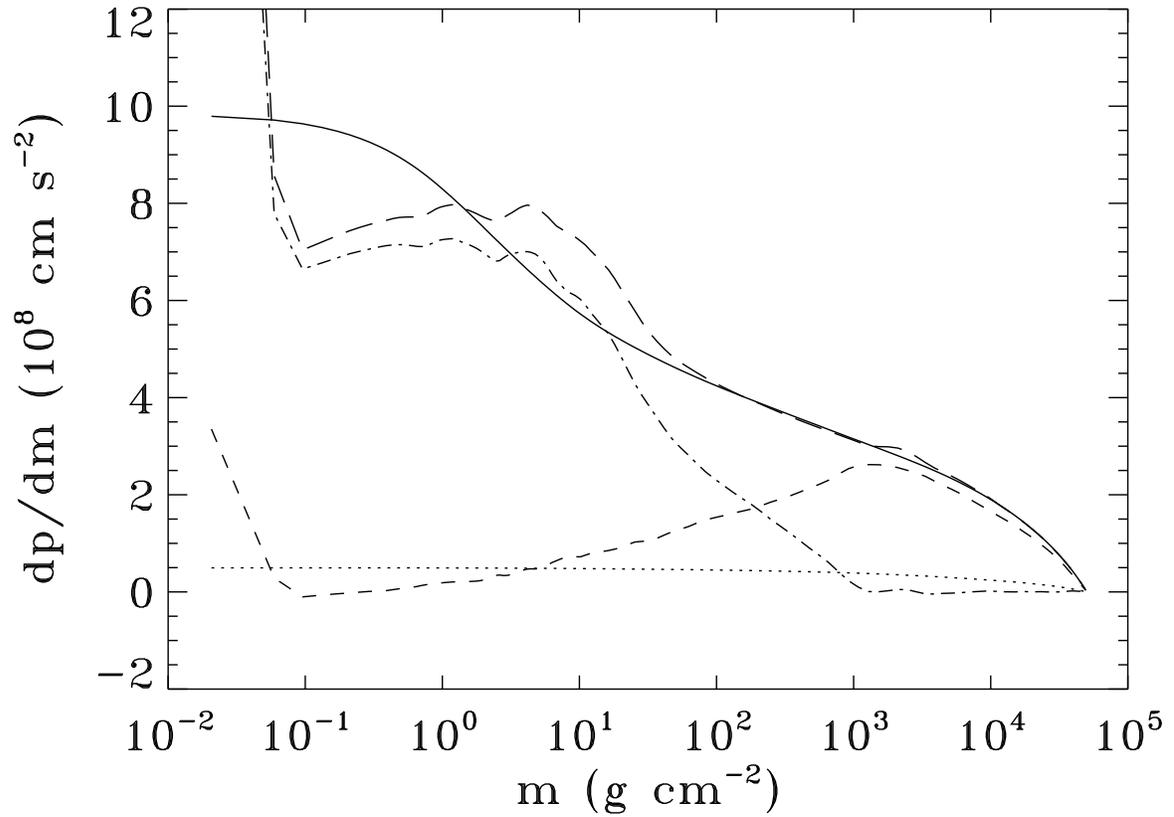}
\caption{Time and horizontally-averaged profiles of vertical acceleration
due to magnetic pressure gradients (dot-dashed), gas pressure gradients
(short-dashed), and radiation pressure (dotted).  The long-dashed curve
shows the sum of these three accelerations, and should be compared to
the magnitude of the gravitational acceleration (solid curve).\label{figdpdm}}
\end{figure}

\begin{figure}
\plotone{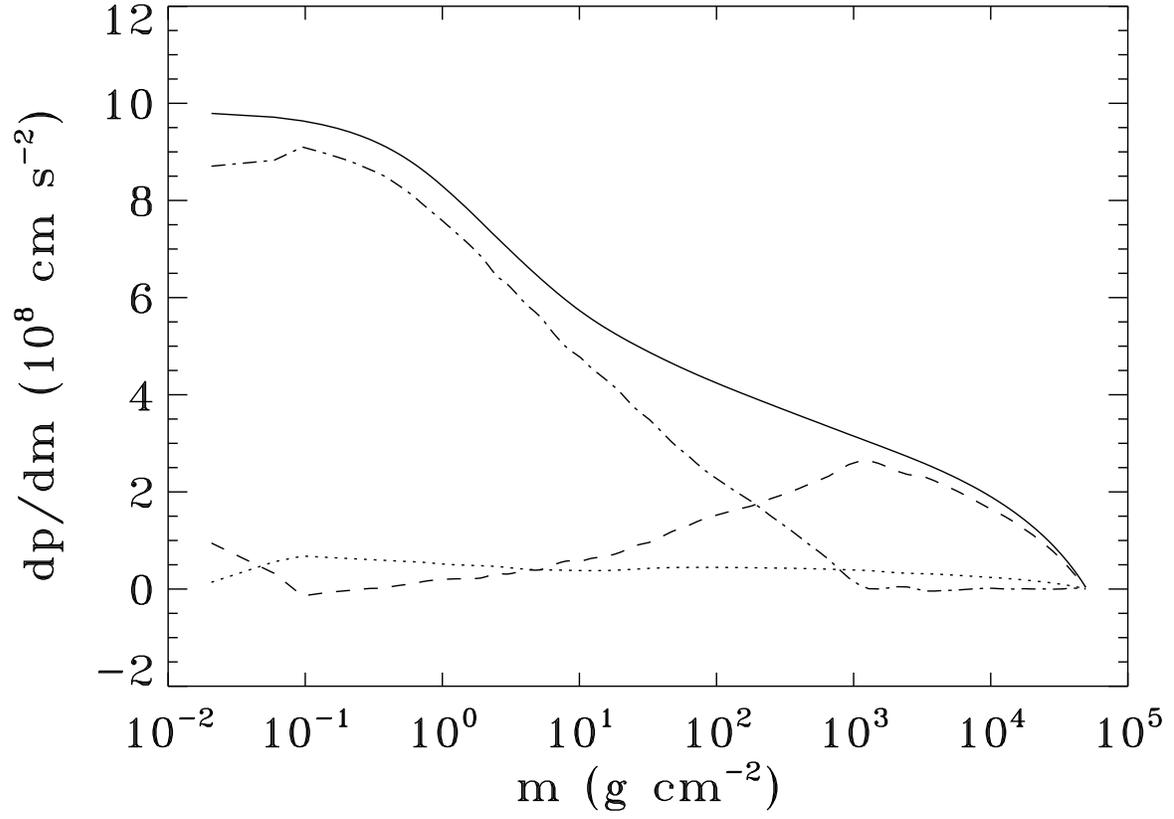}
\caption{Rescaled, fractional contributions of the time and
horizontally-averaged vertical acceleration due to magnetic
pressure gradients (dot-dashed), gas pressure gradients
(short-dashed), and radiation pressure (dotted).  The local gravitational
acceleration is shown by the solid curve.\label{figdpdmfrac}}
\end{figure}

Figure \ref{figdpdm} depicts the average vertical accelerations produced by
magnetic pressure gradients $dp_{\rm mag}/dm\equiv d/dm(B^2/8\pi)$, gas
pressure gradients $dp/dm$, and radiation pressure $\chi F/c$.  Ignoring
magnetic tension forces, the sum of these three should equal the magnitude
of the local vertical gravitational acceleration if the structure was
completely stationary and one-dimensional.  For column masses
$m>10^2$~g~cm$^{-2}$, this is reasonably
accurate, but there are strong deviations from this at lower column densities.
Again, it is beyond the scope of this paper to compute the time-dependent
spectrum from the simulations.  In order to account for the effects of large
magnetic pressure gradients in the upper layers, we determine the fractional
contribution of each acceleration to the total at every location, and then
rescale that
fraction by the local gravity.  The result is depicted in Figure
\ref{figdpdmfrac}.
We then incorporate this profile of $dp_{\rm mag}/dm$ in TLUSTY by mapping
it directly onto the column density grid used in the code.  In order to
avoid the upper vertical boundary effects on the simulation data at low
column densities (see the kinks in Figure \ref{figdpdmfrac} at
$m=0.1$~g~cm$^{-2}$), we
extrapolate $dp_{\rm mag}/dm$ as a constant for $m<0.1$~g~cm$^{-2}$.

\section{Spectra from magnetically supported disks}

\begin{figure}
\plotone{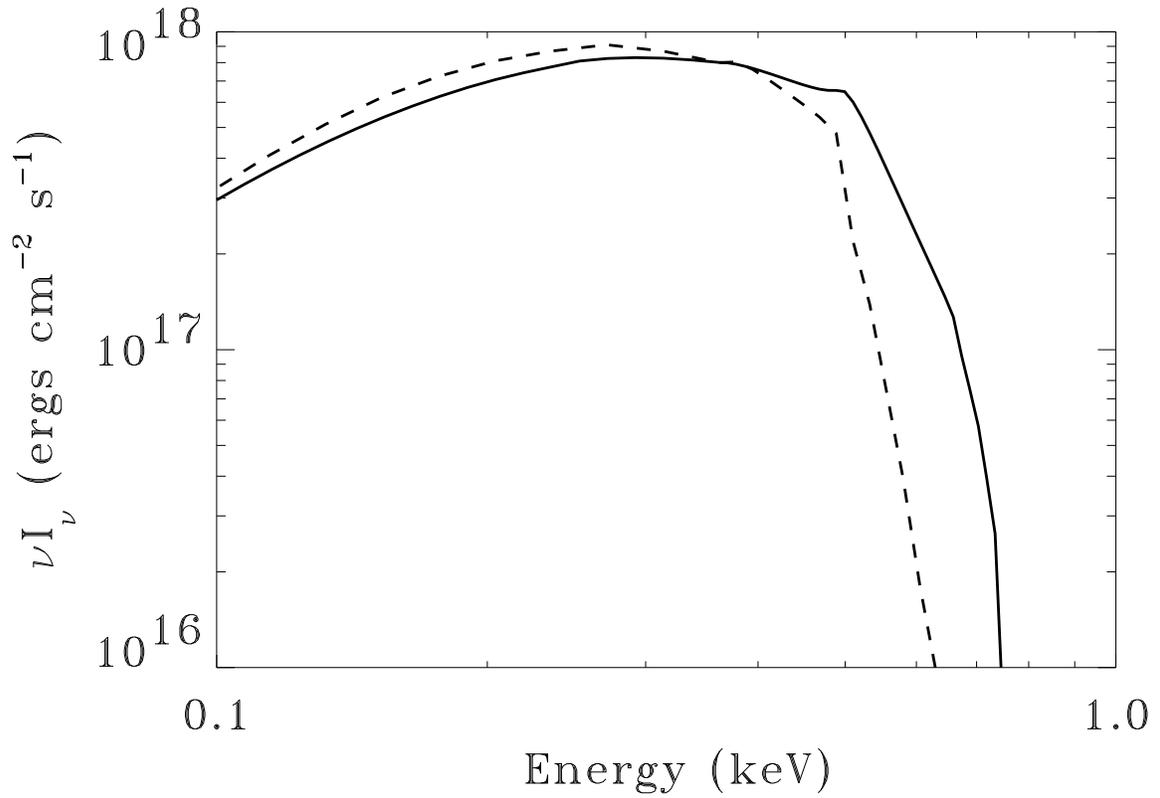}
\caption{Local emergent spectrum in fluid rest frame from the disk annulus
viewed at an inclination angle of $55^\circ$ from the disk normal.  The
dashed curve neglects magnetic pressure support in the vertical hydrostatic
balance, while the solid curve includes it.  Both calculations assume the
dissipation profile of equation (\ref{eqdfdmfit}).  \label{figpmagspec}}
\end{figure}

Figure \ref{figpmagspec} compares the emergent spectrum from the annulus
when magnetic pressure support is or is not included in the vertical
hydrostatic balance.  In contrast to changing the dissipation profile (Figure
\ref{figdissspec}), magnetic pressure support produces a much more
noticeable change, and causes the spectrum to be significantly harder.

\begin{figure}
\plotone{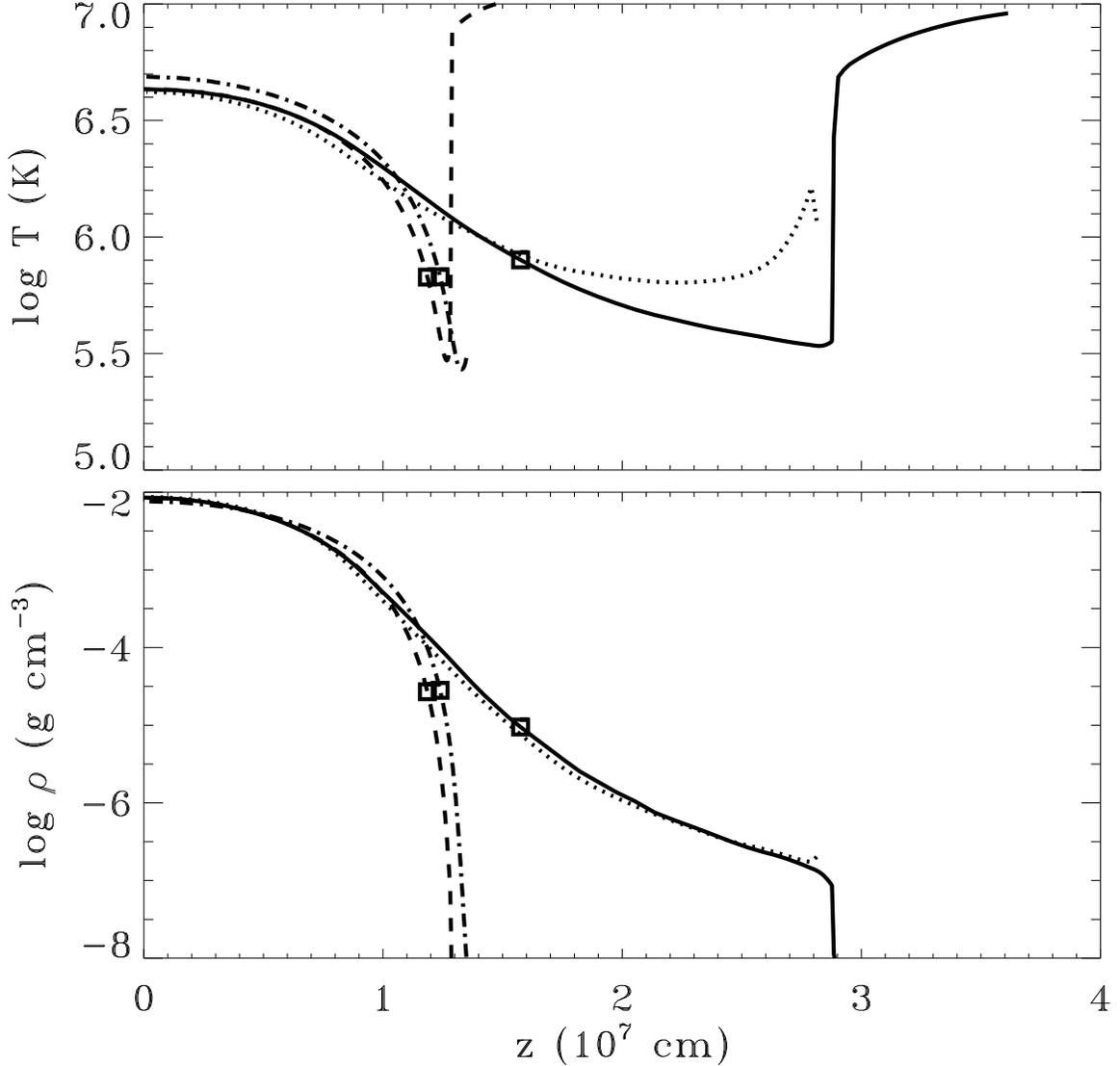}
\caption{Gas temperature (top) and density (bottom) versus height above the
midplane of the annulus.  The solid curve includes the contribution of
magnetic pressure support to the hydrostatic balance, while the dashed
curve neglects it.  Both curves assume the dissipation profile of equation
(\ref{eqdfdmfit}).  The dot-dashed curve is the structure that would be
obtained under standard assumptions:  no magnetic pressure support and a
constant dissipation per unit mass.  Squares mark the position of the
effective photosphere for photon energies of 0.45~keV, which is near the peaks
of the $\nu I_\nu$ spectra shown in Figures \ref{figdissspec} and
\ref{figpmagspec}.  For comparison, the dotted curve
shows the time and space-averaged temperature and density from the
\citet{hir05} simulation.  \label{figpmagvertstruc}}
\end{figure}

\begin{figure}
\plotone{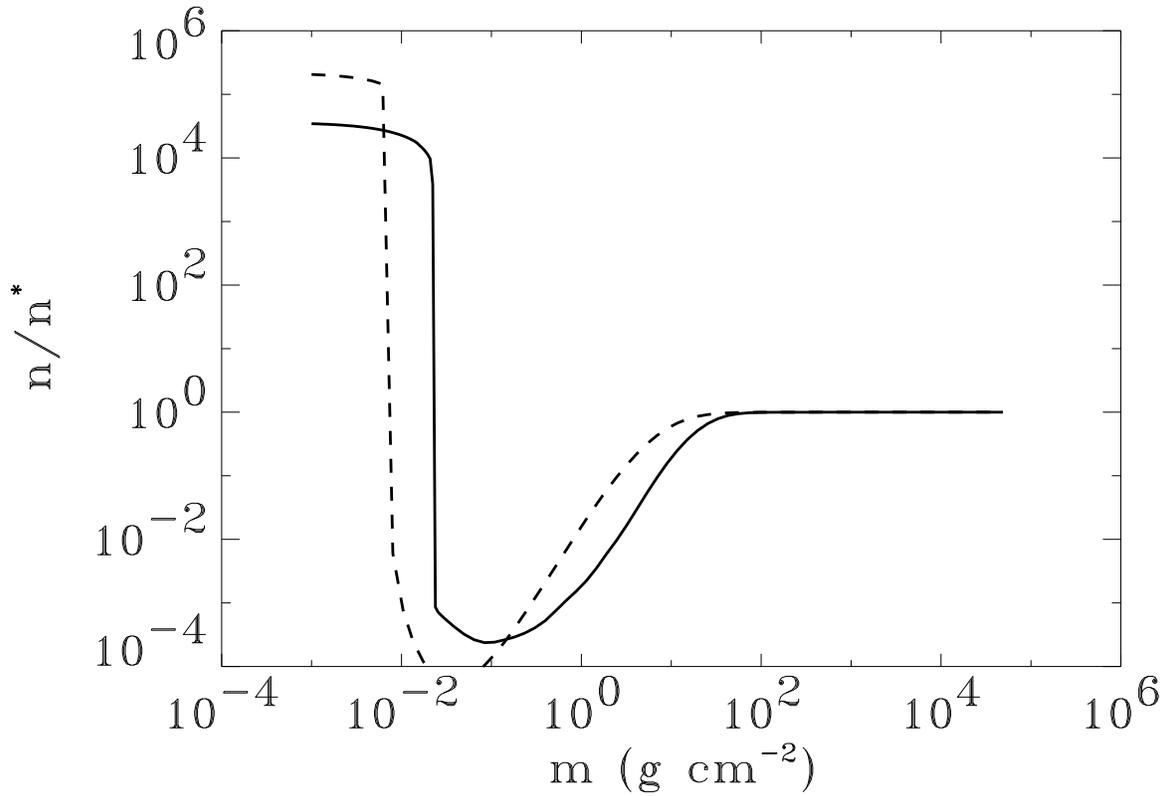}
\caption{Ratio of the actual ground state number density $n$ of CVI to the
ground state number density $n^\star$ that would be present in LTE, as a
function of column density.  The solid curve includes the effects of magnetic
pressure support, while the dashed curve neglects it.  Both curves assume
the numerical dissipation profile of equation (\ref{eqdfdmfit}).  Magnetic
pressure support results in greater departures from LTE at higher column
densities.  \label{figdepcoeff}}
\end{figure}

The reason for this can be seen in Figure \ref{figpmagvertstruc}, which shows
the vertical density and temperature profiles in the two structures, as
well as the locations of the effective photospheres at frequencies
corresponding to the two spectral peaks.  Without magnetic pressure support,
large gas pressure gradients are required to support the disk against the
high gravity at high altitude.  These necessarily require steep density
gradients because the temperature profile (largely set by radiative
equilibrium) is too flat.  Magnetic pressure support relaxes this constraint,
permitting a much more extended density profile.  Because the location of
the effective photosphere is set by an approximately fixed column density, the
larger density scale height in a magnetically supported annulus implies that
the density at the effective photosphere will be lower than would be the
case if magnetic pressure was neglected.  The lower density reduces the
ratio of absorption to scattering opacity, which alone would tend to harden
the spectrum by producing a modified blackbody.  In addition, the lower
density implies that metals are more ionized than they would otherwise
be, and this reduces the strength of absorption edges.  This is illustrated
in detail in Figure \ref{figdepcoeff}, which shows the non-LTE departure
coefficient for the CVI ground state, which is responsible for the absorption
edge feature at $~0.5$~keV in the unmagnetized spectrum.  In the magnetized
atmosphere, non-LTE effects reduce the ground state population of this ion
even further, and drive the edge into emission.  For all these reasons,
we expect atmosphere calculations which include magnetic pressure support
will generically result in harder spectra.  This applies to electrically
conducting accretion disks with MRI turbulence in all astrophysical contexts,
from cataclysmic variables to active galactic nuclei.


Figure \ref{figpmagvertstruc} also compares the time and
horizontally averaged density and temperature structures from the \citet{hir05}
simulation with those computed by TLUSTY.  The density profiles are in
excellent agreement, and even the temperatures compare favorably beneath
the effective photosphere.


Observations of course do not measure the spectrum emitted locally at some
radius in the disk, but the spectrum of the entire disk as carried by photons
that reach detectors at infinity.  To illustate how important
magnetic pressure support might be on an observed system, we calculated the
spectrum of a relativistic accretion disk around a black hole.  This is
a cleaner system than, e.g., a cataclysmic variable where boundary layer
effects need to be included.  However, we warn the reader that
significant uncertainties still remain in our understanding of the nominally
radiation pressure supported inner regions of luminous black hole accretion
disks, uncertainties which we do not address here.

We chose a ten solar mass Schwarzschild hole, accreting at
$3.88\times10^{-8}$~M$_\odot$/yr (giving an Eddington ratio of
$L/L_{\rm Edd}=0.08$).  We assume a standard Shakura-Sunyaev prescription,
where the vertically integrated stress equals the vertically integrated
total pressure (neglecting magnetic fields), times $\alpha=0.01$.

\begin{figure}
\plotone{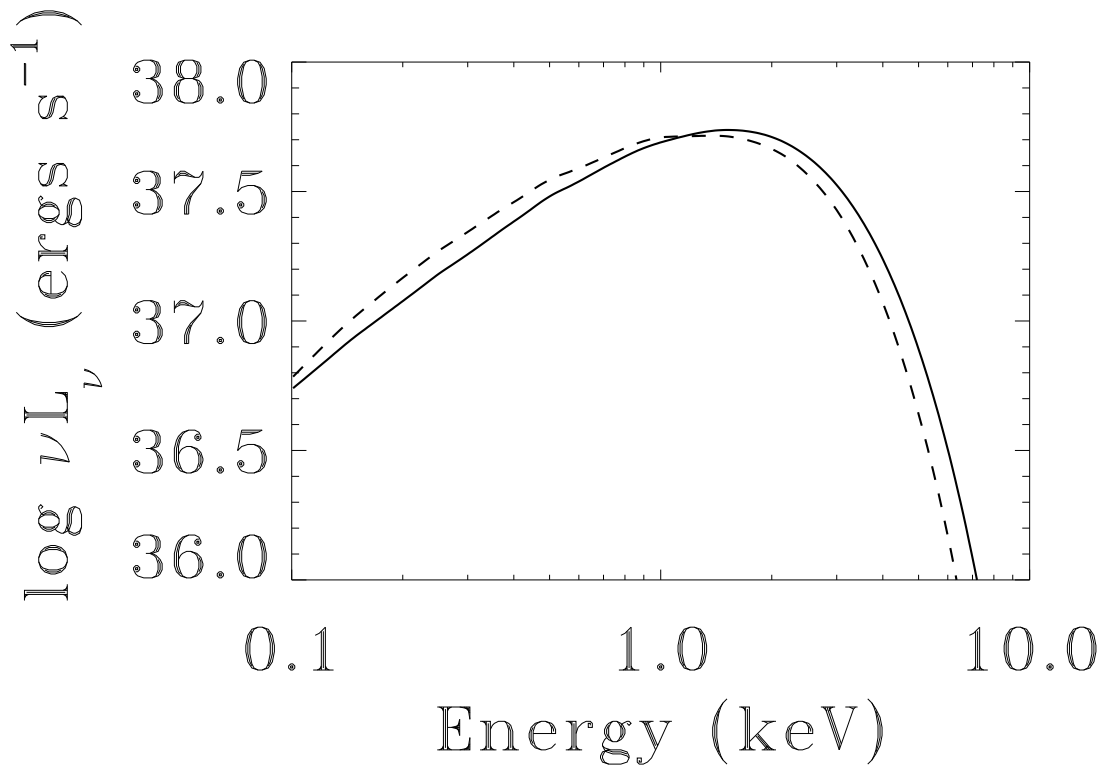}
\caption{Spectrum of full relativistic accretion disk models with
$\alpha=0.01$ and $L/L_{\rm Edd}=0.08$ orbiting a ten solar mass Schwarzschild
hole, as viewed by an observer at infinity at $60^\circ$ from the symmetry
axis.  Both models adopt the vertical dissipation profile given by equation
(\ref{eqdfdmfit}).  The solid curve includes magnetic pressure support, while
the dashed curve neglects it.
\label{wholediskspec}}
\end{figure}

We first calculated the vertical structure of each annulus neglecting magnetic
pressure support but using the dissipation profile of equation
(\ref{eqdfdmfit}).  This determines a midplane pressure $P_0$.  Then combining
this with the surface mass density at the midplane $m_0$ (determined from
the vertically-integrated radial disk structure model), we rescaled our
numerical magnetic pressure acceleration profile by $P_0/m_0$.  We then
used this rescaled magnetic pressure profile to calculate a new model for
the annulus vertical structure.  The resulting disk-integrated spectrum
as seen by an observer at infinity at an inclination angle of $60^\circ$ to
the symmetry axis of the hole is shown in Figure \ref{wholediskspec}.  As
expected, adding magnetic pressure support hardens the spectrum.

In order to quantify the degree of hardening, we fit the spectrum above 0.1~keV
to relativistic, multitemperature blackbody disk models with a constant
color correction factor $f$, defined such that the local specific intensity
is given by an isotropic, color corrected blackbody:
$I_\nu=B_\nu(fT_{\rm eff})/f^4$, where $B_\nu$ is the Planck
function and $T_{\rm eff}$ is the local effective temperature.  We fixed the
mass, Eddington ratio, and disk inclination in these fits, only allowing
$f$ and the overall normalization to vary.  (Allowing the normalization to vary
is necessary as the actual model spectrum is limb-darkened.)  The fitted color
correction factor was 1.74 for the magnetized spectrum, and 1.48 for no
magnetic fields.  We also tried fitting
the spectra above 0.1~keV to relativistic disk atmosphere models around
stellar mass black holes with $\alpha=0.01$ \citep{dav05,dh05}.  In these
fits we again fixed the mass, Eddington ratio, and disk inclination.  We also
fixed the overall normalization, as the relativistic disk atmosphere models
account for limb-darkening.  Only the black hole spin was allowed to vary.
The best fit spin for the unmagnetized
model was $a/M=0.05$, which is consistent with the fact that the modified
dissipation profile in equation (\ref{eqdfdmfit}) does not produce much change
in the overall spectrum.  For the magnetized disk model, however, we found
the best fit spin to be $a/M=0.46$, due to its harder spectrum.  Because
limb darkening has some spin-dependence, we also redid the fits with free
normalization, but still got very similar best-fit spins.

\section{Discussion and conclusions}

The advent of thermodynamically self-consistent simulations of MRI turbulence
in vertically stratified shearing boxes \citep{tur04,hir05} has finally 
opened the door to the construction of spectral models which are based
on the physics of the turbulence itself.  Surprisingly, our investigation
in this paper indicates that the thermodynamics itself, i.e. the vertical
profile of dissipation, has little effect on the emergent spectrum compared
to models based on the the standard assumption of a constant dissipation
rate per unit mass.  This conclusion was also reached in the earlier
investigation of \citet{dav05}.  Unless an annulus in the disk
is only moderately effectively thick, most of the dissipation occurs
deeper than the effective photosphere, and its vertical profile therefore
has little effect on the emergent spectrum.  The reader should bear in
mind, however, that the numerical dissipation profiles, e.g. equation
(\ref{eqdfdmfit}), are based on simulations in which mechanical energy
is simply lost at the grid scale and replaced by internal energy.  More
work needs to be done to investigate whether these profiles are robust to
changes in grid resolution.  In addition, the simulations were for gas
pressure dominated annuli, and large uncertainties still remain to be fully
investigated in the radiation pressure dominated case, which is more relevant
for the innermost annuli of luminous accretion disks around black holes
and neutron stars.

In contrast to the thermodynamics, the actual {\it dynamics}, i.e. the
contribution of magnetic pressure to the vertical support of the disk against
gravity, does produce significant changes in the emergent spectrum when
compared to models based on the standard ad hoc assumptions.  Magnetic pressure
support at high altitude results in a much more extended density scale
height, because gas pressure is not required to support the atmosphere against
the increasing vertical gravity at high altitude.  Because absorption opacity
generally increases with density, a larger density scale height results in
a smaller density at the effective photosphere in order to produce the same
effective optical depth of unity.  As anticipated by \citet{hir05}, this
enhances non-LTE effects in the spectrum because collisional processes in
the plasma are then diminished compared to radiative processes.  Even in LTE,
lower density at the effective photosphere generally implies that the plasma
becomes more ionized, reducing the bound-free opacity.  Electron scattering is
enhanced compared to absorption, driving the emergent radiation spectrum closer
to a modified blackbody.  All of these effects mean that the emergent spectrum
will be generically harder compared to an atmosphere where magnetic support
is neglected.

We emphasize that these spectral implications will apply to all accretion
disks in which the MRI is thought to act, including
active galactic nuclei, cataclysmic variables, and X-ray
binaries.  In the one whole disk model we constructed, around a stellar mass
black hole, we found that the magnitude of this hardening was equivalent to
choosing a color correction factor in a relativistic, multitemperature
blackbody of 1.74, as opposed to 1.48 in the nonmagnetized case.  
This has important implications for recent attempts to measure black hole
spins in X-ray binaries by continuum spectral fitting (e.g. \citealt{sha05}).
Because magnetic pressure support hardens the spectrum, the fitted black hole
spin to the observed continuum will have to compensate by being smaller.

It is conceivable that magnetic pressure support may be exaggerated in the
local shearing box simulations.  Long wavelength Parker instability modes
might be suppressed because they cannot fit inside the box.  This is an issue
which will need further investigation, but we did a preliminary check by
comparing with global, non-radiative, general relativistic MRI simulations
\citep{dev03}.  In the shearing box simulations, magnetic pressure becomes
dominant
when the density falls below 0.03 times the midplane density.  In contrast,
in a global simulation in Schwarzschild geometry, the magnetic pressure
becomes dominant below 0.1 times the local midplane density.  Magnetic support
therefore appears to be even more important in the global simulation which,
however, was non-radiative.

The shearing box simulations show significant time-dependence, horizontal
structure, and
asymmetries above and below the disk midplane, effects which we have ignored
by ruthless averaging.  More sophisticated radiative transfer techniques will
have to be employed to investigate their effect on the emergent spectrum.
Density irregularities in particular are expected to be even more prominent in
the radiation pressure dominated inner regions of black hole accretion disks
\citep{tur03, tur04,tur05}.  Monte Carlo simulations of the photon transfer
through such inhomogeneous structures suggest that the enhanced ratio of
absorption to scattering opacity in the denser regions helps to thermalize
(and therefore soften) the emergent spectrum \citep{dav04}, at least when the
inhomogeneities are treated as static structures.  Whether or not this persists
in a time-dependent calculation, or whether or not the hardening
due to magnetic pressure support turns out to be the more important effect,
remains to be seen.

\acknowledgments

This research was supported by the National Science Foundation
under grant nos. PHY99-07949 and AST 03-07657, and by NASA under
grant no. NAG5-13228.




\clearpage

\clearpage

\clearpage

\clearpage

\clearpage

\clearpage

\clearpage

\end{document}